\documentstyle[12pt]{article}
\begin{document}

\centerline{\Large A.S. Ilyin, K.P. Zybin, A.V. Gurevich. }
\bigskip

\centerline{\Large {\bf Dark matter in galaxies and growth of giant black holes.}}

\begin{abstract}

The connection between dark  matter and giant black holes in the
galactic nuclei is investigated. The joint evolution of dark and
luminous matter in averaged self-consistent gravitational
potential is considered. It is shown that the distribution of dark
matter remains spherically symmetric even in presence of essential
asymmetry of luminous matter in the galaxy.

The kinetic equation describing evolution of dark matter particles
distribution function which takes into account their dynamics in 
gravitational potential and scattering on stars
is derived. It is shown that significant flux of dark matter on a
seed black hole in the center of galaxy should appear. The growth 
law of seed black hole due to absorption of dark
matter is derived. It appears that during the lifetime of galaxy
the seed black hole should grow significantly, up to scales
$10^7-10^8 M_{\odot}$. Brief analysis of observational
data shows that the presented theory is in reasonable agreement
with observations.

\end{abstract}

\section{Introduction}

Recently, intensive development of techniques and new methods in
astrophysical observations had lead to discovery of more than 80
giant black holes in centers of galaxies \cite{1}, \cite{2}. The
masses of the black holes are distributed in the range $2\cdot10^6
- 3\cdot10^9 M_\odot$ ($M_\odot = 2\cdot10^{33}$g is solar mass),
as is shown at Fig.1.  The analysis of observational data revealed
a connection between the giant black hole mass in the center of a
galaxy and mass of the galaxy's bulge.

We remind that bulge is the most dense central spherical component
of a galaxy, containing mainly old stars. The total rotational
velocity of stars in bulge is usually significantly smaller than
velocity dispersion $\sigma$, that is why the form of bulge is 
often close to spherically symmetric.
The density of stars in bulge grows rapidly to the center. 
The size of bulge $r_b \sim1 - 30$kpc, it grows with the galaxy mass.
The masses of giant black holes $M_{bh}$ are
approximately three orders of magnitude smaller than the masses of
host galaxy bulges.

 According to modern knowledge, the main part of matter in the
 Universe is made up by nondissipative nonbaryon dark matter.
The initial density fluctuations of the dark matter in the 
early Universe had grown up forming large-scale nonlinear 
gravitationally bound objects - galactic haloes.
The important feature of these haloes is singular 
distribution of density in their centers \cite{1a}.
Baryon matter trapped by gravitational field of a halo has settled
gradually to its center and formed a galaxy. The presence of
singularity in the center of dark matter distribution allows the
compressing baryon gas to form a primordial black hole \cite{3}
with mass of the order of $10^3 M_{\odot}$.
Note, that in this scenario primordial black hole is situated
exactly in the galaxy dynamical center. Later the seed
black hole grows intensively due to accretion both 
baryonic and dark matter from the bulge. The present paper is
devoted to the analysis of evolution of dark matter and its
absorption by the black hole in the center of a galaxy.

The paper is organized as follows.

In the second section, the
distribution of dark matter in a galaxy and its evolution under
the influence of the averaged self-consistent gravitational field
of baryon matter (bulge) is considered. The problem is solved in
the adiabatic approximation in assumption of spherical symmetry of
baryon matter distribution. 
The role of deviations from spherical
symmetry is analyzed. It is shown that even if the baryon matter
distribution is noticeably asymmetric, the distribution of dark
matter defined by joint action of gravitational fields of baryon
and dark matter remains spherically symmetric.

In the third section, growth of a seed black hole in galactic
center on account of direct capture of dark matter particles
moving in the averaged self-consistent potential  is investigated.
It is shown that the mechanism is of small efficiency in real
conditions and cannot  lead to significant growth of the seed
black hole mass.

In the forth and fifth sections, the evolution of dark matter
distribution function due to the gravitational scattering 
of particles by stars is considered. The scattering lead to
violation of adiabatic approximation. The important
peculiarity of the process is that the mean free path length of
particles is much more than the size of the bulge, hence the
collisions are rare. Taking this into account, in the section 4 we
derive  the collision integral averaged over dark matter particles
oscillations, and write down the kinetic equation describing the
evolution of dark matter distribution function. It is shown that
because of special features of initial distribution function, the
main role is played by diffusion in angular momentum space.

In the fifth section, the solution of the diffusion equation is
obtained, the flux of dark matter on black hole is derived, and
the law of the black hole's growth is ascertained.

In conclusion, we make brief analysis of observational data
showing that  the growth of giant black holes on account of dark
matter absorption should be essential.

On the whole, the presented theory is in reasonable agreement with
existing observational data concerning giant black holes. Further
development both of the observations and the theory and their
detailed comparison are undoubtedly of great interest.

\section{Adiabatic dynamics of gravitating dark matter and baryonic matter.}

Observation show that over distances of the order of the horizon radius our 
Universe is homogeneous, isotropic and expands uniformly. The expansion leads 
to a rapid cooling of matter. A cold gravitating gas is unstable because of the 
action of universal gravitational forces. Growth of the Jeans instability 
creates regions of strong compression with dimensions much smaller than the 
horizon radius $R_H$. This is of decisive importance for the formation 
(in the CDM scenario) of the large-scale structure of matters in the Universe.

The main role in this process is played by dark matter, which manifests itself
only in the gravitational interaction. According to the nowadays knowledge
the baryonic matter in the Universe contribute $\Omega_B<0.05$ and thus the
gravitational processes are determined mostly by nonbaryonic dark matter.
It is supposed as usually that the particles of which dark matter is composed
interact very weakly with one another and with baryonic matter.

 After recombination initial linear density fluctuations $\delta \rho$
 are growing relatively slowly with time $t$  
(due to expansion of Universe)  according to power law: $\delta \rho/ \rho
\propto t^{2/3}$. As soon as the disturbances amplitude reaches unity $\delta
\rho / \rho \propto 1$ the nonlinear processes dominate. Well known
Zeldovich-Harrison spectrum of the initial linear fluctuations is growing with
the wave vector $k$: $\delta \rho(k)\propto k$. Due to this nonlinear stage
$\delta \rho (k_n) / \rho \propto 1$ is reached first for the small $k$.
Nowadays the nonlinear boundary $l_n \propto {k_n}^{-1} \propto 10-100 Mps$.
Thus the following conditions are fulfilled for the dark matter large-scale
structures $l$, which have passed the nonlinear stage of the Jeans instability:
\begin{equation}
\label{2.01} 
\frac{l}{l_f}\to 0,\,\, \frac{l}{R_H} <<1,\,\,
\frac{u}{c}\approx\frac{l}{R_H}<<1.  
\end{equation} 
Here $l_f$ is the free path
of dark matter particles which we consider as noninteracting, $u$ -
characteristic velocities in the inhomogeniety. We see that the nonlinear
stage of Jeans instability is developing in noninteracting, nonrelativistic,
cold gravitating gas.

The general system of
equation, describing the nonlinear gravitational compression of the dark matter
can be presented in a form \cite{1a}:
$$
\frac{\partial
\rho}{\partial t} + \frac{\partial}{\partial \bf r} \left(\rho \bf u\right)=0
$$
\begin{equation}\label{2.02}
\left( \frac{\partial{\bf }}{\partial t} + {\bf u}\frac{\partial {\bf }}{\partial
{\bf r}} \right) {\bf u} +\frac{\partial\psi_d}{\partial {\bf r}}=0
\end{equation}
$$ \Delta
\psi_d =4\pi G\rho
$$
Here $\rho$ is the density, $u$ - velocity of the gas, and
$\psi_d$ - gravitational potential.  
The nonlinear process is
most strongly developing near the density maximums. Let us consider the
vicinity of the one maximum, presenting the initial density distribution $\rho$
in the general form:
\begin{equation}\label{2.2} \rho_{t=0}({\bf r})=\rho_0
\left(1-\frac{x^2}{a^2}-\frac{y^2}{b^2}-\frac{z^2}{c^2}\right)\,,\qquad
\qquad {\bf u}_{t=0}=0
\end{equation}

The solution of hydrodynamical system of equations
(\ref{2.02}), lead to singularity at the moment of time $t=t_k$. After
that moment in a dark matter gas the multifluid motion is developed. It is
accompanied by density oscillation. With time $t\to\infty$ the number of fluids
is infinitely growing.  The density oscillation amplitudes $a_i (t)$ at the
same time is going to zero:  $a_i (t)\propto O(1/t)$. As a result of this
kinetic mixing of fluids {\it the stationary spherically symmetric cluster of dark
matter is established. The main peculiarity of dark matter distribution in this
cluster is the existence of a singularity in its core }
\begin{equation}\label{2.1}
\rho=K r^{-12/7}
\end{equation}
This result was obtained in analytic theory by Gurevich and Zybin \cite{4}. 

It should be noted that numerical calculation could not receive this result
for a long time and showed only the plane structures of the Zeldovich
pancake type. Only the usage of special numerical methods allowed
to make a significant step demonstrating the existence of the stationary
spherical symmetric structures \cite{5}. But the singularity of density
distribution was different from (\ref{2.1}). Further development of
numerical methods allowed to improve this result and move on very close to
the law (\ref{2.1}) \cite{5a}. So nowadays one can say that the numerical
calculation fully confirmed the existence of a stable spherical clusters
with the singularity law (\ref{2.1}). This cluster was called nondissipative
gravitational singularity (NGS).

The main method of analytical solution \cite{4} was based on the solution of 
kinetic equation for dark matter particles using
canonical action variables $I=\left(I_R , m \right)$, where 

\begin{equation}\label{I1}
I_R(E,m)=\frac{\sqrt{2}}{\pi} \int_{r_-}^{r_+} \sqrt{E-\psi_d (r) - \frac{m^2}{2r^2}}dr,
\end{equation}
- radial action, $m$ is the angular momentum, 
$E$ - energy, $r_\pm$-turning points 
(roots of the equation $E-\psi_d (r)-m^2 /2r^2 =0$).
According to \cite{1a} the established distribution function is
\begin{equation}\label{2.3}
f(I_R,m)=f_0 I_R^{1/8} \delta\left(m^2 - {l_0}^2 I_R^2\right),
\end{equation}
where $f_0$ and $l_0$ are constants, determined by the initial distribution
(\ref{2.2}).
In particular the case of spherical symmetric initial density distribution
(\ref{2.2}): $a=b=c$, is degenerate. The spherical symmetry is conserved at all
stage of evolution, including initial compression, hydrodynamical singularity,
kinetic mixing. In this case the angular momentum are always equal to zero,
consequently in (\ref{2.3}) $l_0 = 0$. In the case of weak initial asymmetry: \\
$(a-b)^2 +(a-c)^2 << a^2 $ parameter $l_0$ is small, proportional to
$\epsilon$:
\begin{equation}\label{2.3a}
l_0=0.16\epsilon,\,\,\qquad\qquad
\epsilon=\frac{1}{a}\sqrt{(a-b)^2 + (a-c)^2 - (a-b)(a-c)}
\end{equation}
Accordingly as follows from (\ref{2.3}) angular momentum in this case is also 
small and proportional to parameter $\epsilon$ and to radial action $I_R$.

We will consider now the effect of {\it baryonic}
matter on the distribution of dark matter. Let us take into account, that the
baryonic matter particles lose their energy due to inelastic collisions
accompanied by optic emission. That is why the baryonic matter falls down to
the bottom of the potential wells, formed by the cold dark matter. The luminous
baryonic matter then acts as indicator of the dark matter structure identifying in
particular the position of the centre of NGS. This produces a unique object:
baryonic matter with a halo of dark matter.  Examples of such objects are
galaxies: in this case the evidence for the presence of a dark matter halo is
provided by flat rotation curves \cite{6}. Other
examples of such objects are clusters of galaxies, which contain a trapped hot gas,
again directly confirming the existence of a dark matter halo \cite{7}. 

It should be
pointed out that as the baryonic matter drops to the bottom of potential wells,
its density rises at the central regions of the wells and can become equal to the
density of nonbaryonic matter. That will lead to the considerable distortion of
the canonical distribution of the dark matter (\ref{2.1}).  The main feature
of the problem is that the baryonic matter loses its energy and falls to the
wells bottom slowly - at the timescales comparable with the lifetime of the
Universe. The dark matter particles trapped in the well oscillate many times.
It means that the radial action $I_R$ is integral of motion (adiabatic invariant)
 and distribution function $f_d$ is
conserved and described as previously by formula (\ref{2.3}). But the
dark matter potential in (\ref{I1}) should be changed to the full potential
$\Psi$, determined both by baryonic and dark matter $\Psi=\psi_d +\psi_b$.
Consequently the density $\rho_d$ takes the form:
\begin{equation}\label{2.5}
\rho_d (r,t)=\frac{\sqrt{2}\pi}{r^2}\int_{0}^{\infty}dm^2 \int_{\Omega_E} dE
\frac{f(I_R,m)}{\sqrt{E-\Psi (r,t) - m^2/2r^2}},
\end{equation}
$\Omega_E$ is the region, where
$E-\Psi (r,t)-m^2/2r^2 > 0$.
\begin{equation}\label{I2}
I_R=\frac{\sqrt{2}}{\pi} \int_{r_-}^{r_+} \sqrt{E-\Psi (r,t) - m^2/2r^2}dr.
\end{equation}
Let us consider central part of the object where $M_b>M_d$ and 
$\Psi \approx \psi_b$ . Potential $\psi_b (r)$ is determined by the mass $M_b(r)$ of a 
 baryonic matter. 
$M_b(r)$ is slowly changing with time $t$ due to slowly falling down into the
well of a barionic component. Let us suppose, for example, that $M_b$ has
a power-like dependence on $r$:
\begin{equation}\label{2.6}
M_b (r)\propto r^n,\qquad \psi_b \propto r^{n-1},\qquad n\ge 0.
\end{equation}
From (\ref{2.3}), (\ref{2.5}), (\ref{I2}) and (\ref{2.6}) it follows:
\begin{equation}\label{2.8}
\rho_d = \rho_0 r^{(9n-39)/16}
\end{equation}
Thus we see that in conditions (\ref{2.6}) the power scaling for $\rho_d$ exists
also and just as in the case of purely dark matter (\ref{2.1}),
the scaling law (\ref{2.8}) does not depend on parameters of dark matter initial
distribution $l_0$ and $f_0$. 
For example, if $M_b(r)\propto r^{9/7}$ both
$\rho_d \propto\rho_b \propto r^{-12/7}$. If $\rho_b \propto r^{-2}$, $M_b\propto r$
(isothermal sphere) and
$\rho_d \propto r^{-15/8}$. We see, that dark matter is compressed by
the gravitational action of baryonic gas. For point-like baryonic mass
distribution could be reached maximal compression of a dark matter
\begin{equation}\label{2.9}
\rho_d \propto r^{-39/16}.
\end{equation}
We note that the mass distribution $\rho\propto r^{-m}$ with $m>2$ must lead
to the black hole creation and the problem should be considered in more details
taking this fact into account.

We discussed above the spherically symmetric
distribution of baryonic matter. In real conditions due to rotational motion
baryonic component even in the central parts both of the spiral and elliptic
galaxies can have a noticeable nonspherical components. Can they lead to a
significant discrepancy of dark matter distribution in comparison with the case
of purely spherical symmetric baryonic matter? There are some general theorems 
concerning the behaviour of
 systems close to integrable ones. According to these theorems, the
average action variables $I_R, m, m_z$ remain constant during particles motion 
in weakly asymmetric, slowly
varying potential  \cite{9}. That is why the distribution function (\ref{2.3})
does not change significantly during evolution of the slowly rotating 
baryonic matter \cite{JETF}. 

\section{Black hole growth in the center of the dark matter halo 
(self-consistent approximation).}

The aim of the present section is the investigation of possibility
of seed black holes growth within the bounds of self-consistent, 
collisionless approximation.

The growth of central black hole leads to
decreasing of dark matter particles number in the loss cone. On the other hand, 
it leads to the growth of the loss cone.
As the result of the two competing processes
the flux of a dark matter to black hole can decrease. Let us assume that the
growth of black hole stops at total mass

\begin {equation}\label{4.2}
M_{bh}=M_b+M_d,
\end {equation}
where $M_b$ is an initial baryonic mass of a black hole and $M_d$ is absorbed
dark mass.
The flux of  dark matter on a black hole with mass $M_{bh}$  is composed of
particles with angular momentum

\begin {equation}\label{4.1}
m<m_g = 2cr_g,
\end {equation}
where $r_g = \frac{2M_{bh}}{c^2}$ is the gravitational radius of a black hole.

In spherically symmetric potential $\Psi\left(r,t\right)$ (with arbitrary
time dependence) the angular momentum of particles does not change.
That is why total mass of particles
captured by a black hole is described by initial distribution function $f_i$:
$$
N\left(f_i ; M_{bh} \right) = \int d^3 r d^3 v f_i(\bf r, \bf v)
\theta\left(m_g - m\right),
$$
where

$\theta (x)=1$, if $x>0$, and $\theta(x)=0$, if $x<0$.
The input of a dark matter particles in total mass of black hole is

\begin {equation}\label{4.3}
M_d = N\left(f_i ; M_b + M_d \right)
\end {equation}

Thus the scenario of growth  of black hole mass caused by dark matter particles
flux depends on initial distribution function $f_i$  and black holes baryonic
mass $M_b$  only. If the equation (\ref{4.3}) has the solution
$M_d = M_d\left(f_i ; M_b \right) >0$  the mass of a black hole will not
exceed  $ M_b + M_d \left(f_i ; M_b\right)$. If the equation (\ref{4.3}) has no
solution,  the seed black hole will grow beyond all bounds.

As an example let us consider initial isothermal distribution of a dark matter. 
In this case the distribution function, density and potential have the form:

\begin{equation}\label{4.4}
f_i (E,m) = \frac{\rho_0}{\left(2\pi\sigma_d ^2\right)^{3/2}}e^
{- \frac{E}{\sigma_d ^2}},\quad \rho_0=\frac{\sigma_d^2}{2\pi G},
\end{equation}

\begin{equation}\label{4.5}
\rho(r)=\frac{\rho_0}{r^2},
\end{equation}

\begin{equation}\label{4.6}
\psi_d (r)=4\pi G\rho_0 \ln (r),
\end{equation}
where $\sigma_d$ is velocity dispersion of dark matter particles.
Density of particles with angular momentum smaller than $m_g$
is
$$
\rho_{g} = \frac{\sqrt{2}\pi}{r^2}\int
dE\int dm^2 \frac{f\left(E,m^2\right)}
{\sqrt{E-\psi_d-\frac{m^2}{2r^2}}}\,\,
\theta\left(E-\psi_d-\frac{m^2}{2r^2}\right)\theta(m_g - m)
$$
or, after calculations 
$$
 \rho_g (r)=
  \frac{\rho_0}{r^2}\left( 1- e^{-\frac{m_g ^2}{2\sigma_d ^2 r^2}
  }\right),
$$
Hense total mass of particles with the angular momentum smaller than $m_g$  is
\begin{equation}\label{4.8}
N_g=2\pi \left( \frac{\rho_0}{G}\right)^{1/2} m_g.
\end{equation}
Taking into an account (\ref{4.8}) one can rewrite the
equation (\ref{4.3}) in the form:

\begin{equation}\label{4.9}
\frac{M_d}{M_b}=Q\left(1+\frac{M_d}{M_b}\right),
\end{equation}

where
$$
Q=\frac{8\pi}{c}\left(G\rho_0\right)^{1/2}
$$
The positive solution of the equation (\ref{4.9}) exists only at $Q<1$.
Thus, scenario of growth
of a black hole in the case of isothermal distribution of a
dark matter does not depend on black hole baryonic mass but depends strongly
on the dark matter density determined by factor $\rho_0$.
If $\rho_0$ is anomaly big: $\rho_0>G^{-1}(c/8 \pi)^2$
the initial isothermal distribution is unstable: any small seed
black hole will result in absorption of all dark matter halo.
At $\rho_0<G^{-1}(c/8 \pi)^2$ distribution is steady.
The absorbed dark matter mass appears proportional to baryonic mass and is
determined by expression (\ref{4.9}).

For our Galaxy 
$Q \sim 0.01$ and $M_d$ is less that 1\%, what means, that the accretion 
of dark matter on the initial black hole $M_b$ lead to a very small growth
of its mass only.

Actually, distribution function of dark matter (\ref{2.3}) differs essentially
from isothermal one.
In this case total mass of particles with angular momentum smaller than $m_g$  is determined by
expression
$$
N_g=(2\pi)^3 \int_{0}^{\infty}dI_R \int_{0}^{\infty}dm \int_{-m}^{m}dm_z
f_0 I_R^{1/8}\delta\left(m^2-l_0 ^2 I_R^2 \right) \theta
 \left(m_g - m\right),
$$
hence
\begin{equation}\label{4.12}
N_g=(2\pi)^3 \frac{8}{9}f_0 \left( \frac{m_g}{l_0}\right)^{9/8}.
\end{equation}
The equation (\ref{4.3}) takes the form:
\begin{equation}\label{4.13}
\frac{M_d}{M_b}=Q\left(1+\frac{M_d}{M_b}\right)^{9/8},
\end{equation}
where  $Q=Q'M_b ^{1/8}$, $Q'=(2\pi)^3 \frac{8}{9}f_0 \left( \frac{4G}{l_0 c}\right)^{9/8}$
and $f_0$ can be calculated by expression 

$$
M_H=(2\pi)^3
\int_{0}^{I_{max}}dI_R \int_{0}^{\infty}dm^2 f_0 I_R^{1/8}\delta\left(m^2-l_0 ^2 I_R^2
\right),
$$
where the radial action of particles near the halo's boundary is related to the halo's full 
mass $M_H$ and radius $R_H$ by expression 
$I_{max}\approx\frac{1}{\pi}G^{1/2}M_H ^{1/2} R_H ^{1/2}$,
hense

\begin{equation}\label{4.13g}
f_0=\frac{9}{8}\frac{\pi^{9/8}}{(2\pi)^3}\frac{M_H^{7/16}}{(GR_H)^{9/16}},
\end{equation}

Using this relation, one can show that in our Galaxy the value of parameter $Q$
is again small (of the order of $10^{-2}$), and mass of dark matter captured
by black hole is not significant.

We would like to pay attention that investigation of this problem in previous 
papers \cite{12} results to opposite statement: the initial black hole was
growing without any limits. The reason of this discrepancy is in the fact that
authors did not take into account the changing of dark matter particles 
distribution function
caused by black hole absorption. That is why the loss cone was filled
constantly and the flux on black hole was not exhausted. Considering the
transformation of distribution function one comes to our result.

Thus, dark matter particles cannot cause a noticeable growth
of the black hole because of conservation of their angular momentum in the spherically
symmetric potential.  As it follows from the remark in the end of Section 2,
weak asymmetry of the potential does not change significantly this 
conclusion. The simplest process that can fill the loss cone is 
collisions of dark matter particles with stars. We shall discuss this point 
in next sections.

\section{ Kinetic equation }
Gravitational interaction of dark matter particles and their collisions
with stars is described by
distribution function 
$f({\bf r},{\bf v},t)$, which satisfies the kinetic equation \cite{Fridman}:
\begin{equation}\label{5.1}
\frac{\partial f}{\partial t}+\left\{H_0,f\right\}=St[f] ,
\end{equation}
where $H_0$ is Hamilton function  corresponding to the motion of particles
in the averaged self-consistent potential, $St[f]$  is the collision term.
Gravitational interaction of dark matter particles with individual stars is
described by Coulomb law, hence the collision term may be written in 
Landau form
\cite{13}
\begin{equation}\label{5.2}
St[f]=\frac{\partial}{\partial v_k}W_{kp}\frac{\partial}{\partial v_p}f .
\end{equation}
Here $W_{kp}=2\pi G^2 M_{\odot} \Lambda \int d^3v'w_{kp}F({\bf v'},r), \;
w_{kp}=\left(u^2\delta_{kp}-u_k u_p\right)/u^3$,
$v_k$ are the dark matter velocity  ${\bf v}$ components, $F({\bf v'},r)$ 
is the stars distribution function, ${\bf u}={\bf v'}-{\bf v}$
is the relative velocity of a star and dark matter particle,
$\Lambda \sim 10$  is the gravitational Coulomb logarithm.

In fact, as we have already noticed, the frequency of collisions of particles
with stars is much less than the frequency of their orbital motion.
Thus the kinetic equation (\ref{5.1}) may be significantly simplified.
For this purpose, let us rewrite it in the action-angle variables $I, \phi$:
\begin{equation}\label{5.3}
\frac{\partial f}{\partial t}+\omega_k \frac{\partial f}{\partial
\phi_k}=St_{I,\phi}[f]
\end{equation}
Taking into account that  the initial distribution function (\ref{2.3})
does not depend on angle variables $\phi$, we note that its time variation
is determined by the collision term  (\ref{5.2}) only. However,
the collisions are rare  and $St[f]\propto \nu f$, where $\nu$ is the collision
frequency. Hence we search a solution of the equation  (\ref{5.3}) in the form
$f=f_0+\nu f_1$, where
$\nu f_1$ is a small correction to $f_0$ , and both the terms $f_0$  è $f_1$
depend on "fast" $t_0=t$  and "slow" $t_1=\nu t$ times. In zero approximation
by the parameter $\nu$  we obtain
$$
\frac{\partial f_0}{\partial t_0}+
\omega_k \frac{\partial f_0}{\partial \phi_k}=0
$$
From this equation follows that the main part of distribution
function $f_0$ does not depend on angle variables and "fast" time. The 
first approximation gives
$$
\nu \left(\frac{\partial f_0}{\partial t_1}+ \frac{\partial f_1}{\partial
t_0}+\omega_k \frac{\partial f_1}{\partial \phi_k}\right)=St_{I,\phi}[f_0].
$$

Averaging this equation over the angle variables and "fast" time,
taking into account that the second and third terms become zero, 
we obtain equation
\begin{equation}\label{5.7}
\frac{\partial f(I,t)}{\partial t}=\overline{St}[f(I,t)]
\end{equation}
(Here we omit the index   "$0$" in distribution function and returned to 
usual time  $t$.)

So, under assumptions on collision frequency  $\nu$ made above,
it is possible to find the distribution depending on action variables
only, as a solution of kinetic equation  (\ref{5.7}) with time-averaged
collision term: 
\begin{equation}\label{5.8}
\overline{St}[f]=\frac{1}{(2\pi)^3}\int d^3\phi St_{I,\phi}[f]
\end{equation}

The time-independent kinetic equation in the averaged form (\ref{5.7})
describing the distribution of stars in the vicinity of a giant black hole was
first studied in \cite{14}, \cite{15} and \cite{shap}. A purely Coulomb 
potential field was considered in these works.

Our aim is to examine time-dependent solutions of the kinetic equation 
(\ref{5.7}) describing dynamics of particles in arbitrary central 
symmetric field. First, it is necessary to obtain the averaged 
collision integral (\ref{5.8}). To do this we will follow the technique, developed in \cite{16}. 

Let us consider
tensor differential form  (\ref{5.2}) formally in six-dimensional phase space
$X=({\bf v},{\bf x})$, assuming  coordinate components of the tensor $W$ 
being zero:
$$
\frac{\partial}{\partial v_k}W_{kp}\frac{\partial}{\partial v_k}=
\frac{\partial}{\partial X_\mu}W_{\mu \nu}\frac{\partial}{\partial X_\nu};
\qquad k,p=1,2,3; \quad \mu,\nu=1,\ldots,6
$$
In this six-dimensional space we make canonical transformation to the
action-angle variables
$$
X_\mu \mapsto Y_{\mu'}; \; Y=\left\{I_R,m,m_z;\phi_1,\phi_2,\phi_3\right\} .
$$
The differential form then takes the form
$$
\frac{\partial}{\partial X_{\mu}}W_{\mu\nu}\frac{\partial}{\partial X_{\nu}}=
\frac{1}{\sqrt{g}}
\frac{\partial}{\partial Y_{\mu'}}\sqrt{g} R_{\mu' \nu'}\frac{\partial}
{\partial Y_{\nu'}}, \qquad R_{\mu' \nu'}=\frac{\partial Y_{\mu'}}
{\partial X_{\mu}}\frac{\partial Y_{\nu'}}{\partial X_{\nu}}W_{\mu,\nu}
$$
where $\sqrt{g}$ is the Jacobian. However, Jacobian of canonical transformation
is unity, hence the differential form in new variables preserves in the form:
$$
\frac{\partial}{\partial X_{\mu}}W_{\mu\nu}\frac{\partial}{\partial X_{\nu}}=
\frac{\partial}{\partial Y_{\mu'}} R_{\mu'\nu'}\frac{\partial}{\partial
Y_{\nu'}}.
$$
We search for the solution of kinetic equation (\ref{5.7})
which does not depends on angle variables  $\phi$, hence
 $\frac{\partial}{\partial Y}R\frac{\partial}{\partial \phi}=0$.
After averaging  over $\phi$ terms
 $\frac{\partial}{\partial \phi}R\frac{\partial}{\partial I}$ 
is vanished, and the expression for averaged collision integral becomes again
 three-dimensional: 
\begin{equation}\label{5.12}
\overline{St}[f]=\frac{\partial}{\partial
I_{k'}}\overline{R}_{k'p'}\frac{\partial}
{\partial I_{p'}}f,
\end{equation}
where
\begin{equation}\label{5.13}
\overline{R}_{k'p'}=\frac{1}{(2\pi)^3}\int d^3\phi\frac{\partial
I_{k'}}{\partial v_k}
\frac{\partial I_{p'}}{\partial v_p}W_{kp}.
\end{equation}

To give concrete expression to the differential
(\ref{5.12}), one should know the distribution function of 
stars $F$. The observations of stellar dynamics in the bulge indicate that
in the first approximation the distribution function of stars could be assumed
isotropic, i.e. depending on energy  $E'=\frac{v'^2}{2}+\Psi(r)$ only.
For simplicity we also assume that
the dependence of stellar distribution function on energy is power-like:
\begin{equation}\label{5.14}
F(v',r)=F_0 E'^{-\beta}
\end{equation}

One can show that the potential produced by stars distributed in accordance
with  (\ref{5.14}) is also power function of  $r$:
\begin{equation}\label{5.15}
\Psi(r)=\Psi_0r^\alpha,\alpha=4/(2\beta-1)
\end{equation}

We note that both distribution function (\ref{5.14}) and potential
(\ref{5.15}) can be determined if the velocity dispersion of  stars 
$\sigma=\sqrt{\left<v_k^2\right>}=\sqrt{\left<v^2/3\right>}$ is known.
The dispersion could be choosen as a function of distance to the galaxy center: 
\begin{equation}\label{5.16}
\sigma(r)=\sigma_0r^{\alpha/2}.
\end{equation}
The distribution function  (\ref{5.14}) and the potential (\ref{5.15})
are uniquely dependent functions of
the velocity dispersion, and hence the parameters $\sigma_0, \alpha$ only:
$$
\Psi_0=3\sigma_0^2\frac{\int_{0}^{\infty}x^2\left(1+x^2/2\right)^{-\beta}dx}
{\int_{0}^{\infty}x^4\left(1+x^2/2\right)^{-\beta}dx}, \qquad
F_0=\frac{\alpha(1+\alpha)}{(4\pi)^2G\int_{0}^{\infty}x^2\left(1+x^2/2\right)^
{-\beta}dx}\Psi_0^{2/\alpha}
$$

Note that in the case $\alpha=0$ the relations formally lose their mathematical meaning.
However, one can show that when $\alpha \to 0$ the stellar distribution 
function becomes isothermal:

$$
F(E') = \frac{\rho_0}{\left(2\pi\sigma^2\right)^{3/2}}e^
{- \frac{E'}{\sigma^2}},\quad \rho_0=\frac{\sigma^2}{2\pi G},
$$

$$
\Psi(r)=2\sigma^2 \ln (r),
$$
and the velocity $\sigma$
dispersion does not depend on distance. As a rule, the parameter   $\alpha$
in a bulge is rather small.

Calculations show that for isotropic stellar  distribution function  $F(E')$
tensor $W_{kp}$ in  (\ref{5.2}) takes the form
\begin{equation}\label{5.13a}
W_{kp}=A(E,r)\delta_{kp}-B(E,r)\frac{v_k v_p}{v^2},
\end{equation}
where
$$
A=\frac{16\pi^2}{3}G^2 M_\odot \Lambda
\int_{\Psi(r)}^{\infty}dE'F(E')
\left\{
\begin{array}{rcl}
1&,&E<E'\\
\frac{3}{2}\frac{v'}{v}\left(1-\frac{v'^2}{3v^2}\right)&,&E>E'\\
\end{array}
\right.
$$
$$
A-B=\frac{16\pi^2}{3}G^2 M_\odot \Lambda
\int_{\Psi(r)}^{\infty}dE'F(E')
\left\{
\begin{array}{rcl}
1&,&E<E'\\
\frac{v'^3}{v^3}&,&E>E'\\
\end{array}
\right.
$$

We notes, that the radial action $I_R$ of particles
moving in the potential (\ref{5.15}) could be with good accuracy approximated by
the expression
\begin{equation}\label{5.19}
I_R(E,m)\approx J(E)-b_\alpha m,
\end{equation}
where $b_\alpha \sim 1$ is some positive constant,
$$
J(E)=\frac{\sqrt{2}\int_{0}^{1}\left(1-x^\alpha\right)^{1/2}dx}{\pi\Psi_0^{1/\alpha}}
E^{\frac{1}{\alpha}+\frac{1}{2}}, \quad \alpha>0,
$$
$$
J(E)= \frac{\sigma}{\sqrt{\pi}} \;
e^{\frac{E}{2\sigma^2}}, \quad \alpha=0.
$$
(We notice that in the case of Coulomb or oscillator potentials the equality
(\ref{5.19}) becomes exact and the constant $b=1$. In the case of isothermal
stellar distribution $b \approx 0.6$)

The initial function (\ref{2.3}) does not depend on the variable $m_z$,
hence one can search for solution of the equation  (\ref{5.12}) as a function
of   $I_R$  and $m$. Calculating the coefficients of quadratic form
(\ref{5.13}) and taking into account (\ref{5.13a}), we rewrite the 
collision term in a form  
\begin{equation}\label{5.23}
\overline{St}[f]=\frac{1}{m}\frac{\partial}{\partial m}m
\left(\overline{R}_{22} \frac{\partial f}{\partial m}
+\overline{R}_{12} \frac{\partial f}{\partial I_R}\right)
+\frac{\partial}{\partial I_R}
\left(\overline{R}_{12} \frac{\partial f}{\partial m}
+\overline{R}_{11} \frac{\partial f}{\partial I_R}\right),
\end{equation}
where
$$
\overline{R}_{11}=p-2b_{\alpha}q+b_{\alpha}^2s,
$$
$$
\overline{R}_{12}=q-b_{\alpha}s,\qquad \overline{R}_{22}=s,
$$
$$
p=\left(\frac{dJ}{dE}\right)^2
\left<(A-B)v^2\right>_\phi, \;
q=\left(\frac{dJ}{dE}\right)
\left<(A-B)m\right>_\phi,
$$
$$
s=\left<Ar^2-\frac{B}{v^2}m^2\right>_\phi,
$$
where
$<\ldots>_\phi=\frac{2}{T(E,m)}\int_{r_-}^{r_+}\frac{dr}{v_r}(\ldots),$
and $T=2\int_{r_-}^{r_+} \frac{dr}{v_r}$ is the period of dark matters 
partical radial oscillations,
 $v_r=\sqrt{2}\left(E-\Psi(r)-\frac{m^2}{2r^2}\right)^{1/2}$ is
radial velocity. 
Taking into account that angular momentum of dark matter particles
are small, in the first approximation we shall calculate the 
coefficients  $\overline{R}_{ab}(I_R,m)$ in  (\ref{5.23})  at  $m=0$.
Further, from the initial distribution function  (\ref{2.3}) one can see
that for sufficiently large time, as far as the distribution function has
sharp maximum at $m=l_0I_R,\; l_0 \ll 1$, one can neglect the diffusion by $I_R$
compared to diffusion by $m$.  Summarizing, we rewrite the kinetic equation
(\ref{5.7}) as the diffusion equation depending on $I_R$ only parametrically:
\begin{equation}\label{5.26}
\frac{\partial f(I_R,m,t)}{\partial t}=R(I_R)\frac{1}{m}\frac{\partial}{\partial m}
m\frac{\partial}{\partial m}f(I_R,m,t)
\end{equation}
Here the diffusion coefficient
$R=\overline{R}_{22}|_{m=0}=\left<Ar^2 \right>_{\phi}$ 
calculated with account of   (\ref{5.16}), takes the form
\begin{equation}\label{5.27}
R(I_R)=0.46GM_{\odot} \Lambda \sigma_0^{\frac{2}{2+\alpha}}I_R^{\frac{\alpha}{2+\alpha}}
\end{equation}

We note that in the case of isothermal distribution of stars in the bulge
 $\alpha=0$, and the  diffusion coefficient $R$ does not depend on $I_R$.
We emphasize that the fact that $I_R$ comes into the final equation only
as parameter is conditioned by the particular form of initial distribution
function  (\ref{2.3}) containing small parameter $l_0$.

\section{Flow of dark matter on a black hole.}

As a result of direct capture of particles in the loss cone of seed black
hole, the distribution function of dark matter particles takes the form
\begin{equation}\label{6.1}
f(I_R,m,0)=f_0I_R^{1/8}\delta\left(m^2-l_0^2I_R^2\right)\theta\left(m-m_g\right).
\end{equation}
It differs from  (\ref{2.3}) by  factor $\theta\left(m-m_g\right)$.

The diffusion of dark matter particles in angular momentum space, their drift
into the loss cone  (\ref{4.1}) and further capture by black hole can be 
described by the diffusion equation (\ref{5.26}) with initial condition 
(\ref{6.1}) and boundary condition
\begin{equation}\label{6.2}
f|_{m=m_g}=0.
\end{equation}
The solution of the diffusion equation can be presented in a form
\begin{equation}\label{6.3}
f(I_R,m,t)=\int_{m_g}^{\infty}dm_1G(I_R,m,m_1,t)f(I_R,m,0)
\end{equation}
where $G$ is  the Green function of the boundary problem 
(\ref{5.26})-(\ref{6.2})

$$
G=\int_{0}^{\infty}d\lambda m_1Z_\lambda \left(m_1,m_g\right)
Z_\lambda \left(m,m_g\right)e^{-\lambda R(I_R)t},
$$
$Z$ is the orthogonal system of fundamental functions of the boundary problem 
(\ref{5.26})-(\ref{6.2})
$$
Z_\lambda\left(m,m_g\right)=\frac{J_0\left(\sqrt{\lambda}m_g\right)
N_0\left(\sqrt{\lambda}m\right)-N_0\left(\sqrt{\lambda}m_g\right)
J_0\left(\sqrt{\lambda}m\right)}{\left(J_0^2\left(\sqrt{\lambda}m_g\right)
+N_0^2\left(\sqrt{\lambda}m_g\right)\right)^{1/2}},
$$
$J_0,N_0$ are Bessel and Neumann functions of zeroth order.

Let us now calculate the flux of dark matter on a black hole. Let $D$ be
the area in the phase space where
 $I_R\ge 0,m\ge m_g,-m\le m_z\le m$.
 The whole mass of dark matter is  $N(t)=\int_{D}d^3I d^3 \phi f(I_R,m,t)$. 
 Now, from the condition of mass conservation it follows that the flux 
 through the boundary is   $S=-dN/dt$.
 Taking into account   (\ref{5.2}) we obtain
$$
S(t)=-\int_{D}d^3I d^3\phi \frac{\partial}{\partial I_k}\overline{R}_{kp}
\frac{\partial}{\partial I_p}f(I_R,m,t).
$$

Let us transform now the integral over volume into an integral over surface
according to Stokes formula. Taking into account that the only flow of
matter is through the surface  $m=m_g$ , we find that
\begin{equation}\label{6.7}
S(t)=2(2\pi)^3\int_{0}^{\infty}dI_R R(I_R) m_g 
\left. \frac{\partial}{\partial
m}f(I_R,m,t)\right|_{m=m_g}.
\end{equation}
Substituting the solution of diffusion equation (\ref{6.3}) into (\ref{6.7}),
we obtain
$$
S(t)=2\frac{(2\pi)^3}{\pi}f_0 R_{\alpha}\int_{0}^{\infty}dI_R I_R^{\gamma+\frac{1}{8}}\int_{0}^{\infty}d\lambda e^{-\lambda R_\alpha I_R^{\gamma}t}
\frac{Z_{\lambda}\left(l_0I_R,m_g\right)}{\left(J_0^2\left(\sqrt{\lambda}m_g\right)+
N_0^2\left(\sqrt{\lambda}m_g\right)\right)^{1/2}},
$$
where 
\begin{equation}\label{6.7a}
\gamma=\alpha/(2+\alpha), R_{\alpha}=0.46\Lambda
GM_{\odot}\sigma_0^{\frac{2}{2+\alpha}}.
\end{equation}
It is convenient to introduce a normalized time $T=l_0^{-\gamma}R_{\alpha}t$ 
and make a change of variables
$$
\lambda\mapsto\eta=\lambda T^{\frac{2}{2-\gamma}}, \
I_R\mapsto y=l_0 I_R  T^{-\frac{1}{2-\gamma}}, \
m_g\mapsto x=m_g T^{-\frac{1}{2-\gamma}}
$$
The expression for the flux takes then a form
\begin{equation}\label{6.7b}
S(t)=2\frac{(2\pi)^3}{\pi}\frac{f_0}{l_0^{\frac{9}{8}+\gamma}}
\frac{R_{\alpha}}{T^{\zeta}}\Phi_{\gamma}(x),
\end{equation}
$$
\zeta=\frac{7-8\gamma}{16-8\gamma},\qquad
\Phi_{\gamma}(x)=\int_{x}^{\infty}dy H_{\gamma}(x,y),
$$
$$
H_{\gamma}(x,y)=y^{\frac{1}{8}+\gamma}
\int_{0}^{\infty}d\eta e^{-\eta y^{\gamma}}
\frac{J_0\left(\sqrt{\eta}x\right)
N_0\left(\sqrt{\eta}y\right)-N_0\left(\sqrt{\eta}x\right)
J_0\left(\sqrt{\eta}y\right)}{J_0^2\left(\sqrt{\eta}x\right)
+N_0^2\left(\sqrt{\eta}x\right)}.
$$
One can show that   $H_{\gamma}(x,y)$ is finite positive function, it is 
equal to zero at  $y=x$  and exponentially small at $y>x+4$.
From this follows that at the moment $t$ the flow of dark matter to 
the black hole is made up by particles with radial action range in an
interval
\begin{equation}\label{6.7b}
\frac{m_g}{l_0}<I_R<\frac{m_g}{l_0}+\frac{4}{l_0}
T^{1/(2-\gamma)}.
\end{equation}
Thus, the area from where the dark matter flows into 
black hole, grows with time as  $t^{1/(2-\gamma)}$.
For function $\Phi$ in expression (\ref{6.7b}), there is an estimate
$\Phi_{\gamma}(x)\approx x^{\frac{1}{8}+\frac{\gamma}{2}}.$
Finally, we find the expression for the flux of dark matter onto a black
hole through the boundary of the loss cone $m=m_g$: 
\begin{equation}\label{6.8b}
S(t)=2\frac{(2\pi)^3}{\pi}\frac{f_0}{l_0^{\frac{9}{8}+\gamma}}R_{\alpha}^{1/2}
\frac{m_g^{\frac{1}{8}+\frac{\gamma}{2}}}{t^{1/2}}.
\end{equation}

The expression (\ref{6.8b}) has been obtained in assumption that 
the boundary of the loss cone  $m_g$ does not depend on time.
In reality, the boundary moves in accordance with the whole black hole mass.
However, from  (\ref{6.8b}) follows that the flux depends only weakly on
the value  $m_g$. It means, that the growing of $m_g$ is a quasistationary
process and one can use the expression (\ref{6.8b}),
assuming $t$ as a parameter: $m_g(t)=4GM_{bh}(t)/c$.
Hence, the  growth of the black hole is described by
$$
\frac{dM_{bh}}{dt}=2\frac{(2\pi)^3}{\pi}\left(\frac{4G}{c}\right)^
{\frac{1}{8}+\frac{\gamma}{2}}\frac{f_0}{l_0^{\frac{9}{8}+\gamma}}
R_{\alpha}^{1/2}\frac{M_{bh}^{\frac{1}{8}+\frac{\gamma}{2}}}{t^{1/2}}.
$$
Assuming the seed black hole mass small, we obtain the solution of this 
equation
\begin{equation}\label{6.10}
M_{bh}=C t^{\frac{4}{7-4\gamma}},
\end{equation}
$$
C=\left(\left(\frac
{7-4\gamma}{2}\right)\frac{(2\pi)^3}{\pi}\left(\frac{4G}{c}\right)^
{\frac{1}{8}+\frac{\gamma}{2}}\frac{f_0}{l_0^{\frac{9}{8}+\gamma}}
 R_{\alpha}^{1/2}\right)^{\frac{8}{7-4\gamma}}.
$$
To estimate the parameter $f_0$ one can use the relation (\ref{4.13g}).
Thus, the growth of black hole at the expense of absorption of 
dark matter scattered by stars follows a power law
$M_{bh}\propto t^a,\; a\approx4/7$.

\section{Conclusion.}

In conclusion, we compare roughly the results of the theory presented
above with the observational data. Fig.2 represents the dependence
of stellar velocity dispersion on distance to the galactic center
for our Galaxy and galaxies M31, NGC4258 \cite{17}. One can see
that the distribution of stars in bulges  of M31  and NGC4258 is
close to isothermal, i.e. up to the  area of action of black hole
($r<(4-7)$pc) the velocity dispersion is almost constant, it does
not depend on $r$: $\sigma=\sigma_0 \sim 200$km/s.
In the case of isothermal distribution of stars in the bulge we can 
rewrite the expression (\ref{6.10}):

\begin{equation}\label{7.1}
M_{bh}=8 \cdot 10^7 M_{\odot}
\left(\frac{l_0}{0.1}\right)^{-\frac{9}{7}}
\left(\frac{M_H}{10^{12}M_\odot}\right)^{\frac{1}{2}}
\left(\frac{R_H}{100kps}\right)^{-\frac{9}{14}}
\left(\frac{\sigma}{200km/s}\right)^{\frac{4}{7}}
\left(\frac{t}{3 \cdot 10^{17}s}\right)^{\frac{4}{7}}
\end{equation}

Let us suppose that parameter $l_0$ of the dark matter
haloes around these galaxies is close to average value $<l_0>=0.1$.
Implying, that masses $M_H$ and sizes $R_H$ of haloes are roughly
equal and constitute $10^{12}M_\odot$ and $100$kpc
respectively we obtain from (\ref{7.1}) the mass of the black hole
$M_{bh}\approx 8 \cdot 10^{7}M_\odot$ (We assume, that the age $t$ of black
hole is comparable with the age of the Universe $t\approx 3\cdot
10^{17}$s,). Thus, theoretical
predictions of masses of black holes in these galaxies are close
enough to the observed values:
$M_{bh}=(2.0-8.5)\cdot 10^7 M_{\odot} $ for M31 and
$M_{bh}=(3.8-4.0)\cdot 10^7 M_{\odot} $, for NGC4258.

The central part of our Galaxy's bulge does not satisfy the
assumption of isothermal distribution of stars at $r\leq 100$pc.
At the same time, as it is clear from (\ref{6.7b}), the dark
matter accumulates mainly from the central region. The dependence
of stellar velocity dispersion on distance is shown at Fig. 3
\cite{18}. One can see that at the scales from $10$pc to $100$pc
the dispersion can be roughly represented as
$\sigma=60$km/s$\left(\frac{r}{10nc}\right)^{0.3}$. From
(\ref{5.16}) then follows $\alpha\approx0.6$. In the case, the
formula (\ref{6.10}) gives for the black hole mass a value $2\cdot
10^7M_\odot$, which is much more than observed value $M_{bh}
\approx 2.6 \cdot 10^6M_\odot$. To obtain more precise estimation
of the black hole mass, more information on the stellar and dark
matter distribution functions and {\it their evolution} is needed.
Also, the contribution of the black hole into the joint
gravitational potential may be important. However, we see that
even the rough calculation of absorption of dark matter only may
give a reasonable estimate of black hole mass. This estimate is
consistent with observed masses of {\it most} giant black holes (see
Fig.1). The fact may be considered as an additional prove of the
general theory of the role of CDM in the large-scale structuring
of the Universe \cite{4}.

Note also that the kinetic theory developed in Section 5 allows to
describe the absorption of luminous (baryon) matter as well as
dark matter. More precisely, it is possible to take into account the
capture of stars by black hole because of their
gravitational scattering. This process is especially important for
giant black holes having mass of the order $M_{bh} \sim 10^9 M_\odot$ and
active galactic nuclei.

Another important process described by the kinetic equation
(\ref{5.7}) is the possibility of diminishing of amount of dark
matter in  bulge. The change of dark matter particles' energy as
a result of gravitational scattering by stars leads to their
"heating" and pushing them out of bulge -- the phenomenon
analogous to Fermi acceleration. Therefore, the density of dark
matter compared to that of luminous matter decreases.
Qualitatively this corresponds to the results of recent
observations discussed in the literature \cite{19}, \cite{20}.

Further development of the consistent kinetic theory discussed
above and its detailed comparison with observational data would be
undoubtedly useful for understanding of main physical processes in
the galactic nuclei.

The authors are grateful to V.A. Sirota and M.I. Zelnikov for
numerous and useful discussions.

\end{document}